\documentclass[pra, aps, twocolumn, floatfix,superscriptaddress,amsmath]{revtex4-1}
\usepackage{graphicx}
\usepackage{graphics}
\usepackage{mathtools}
\usepackage{float}
\usepackage{amssymb}
\usepackage{color}
\def \bk{{\bf k}}
\def \br{{\bf r}}

\def \mB{\mathrm{B}}
\def \md{\mathrm{d}}
\def \ms{\mathrm{s}}
\def \mm{\mathrm{m}}
\def \meq{\mathrm{eq}}
\def \mw{\mathrm{w}}

\def \mmax{\mathrm{max}}

\def \mdet{\mathrm{det}}

\def \mBKT{\mathrm{BKT}}

\def \cD{{\cal{D}}}

\def \la{{\langle}}
\def \ra{{\rangle}}

\begin{document}

\title{Superfluidity and relaxation dynamics of a laser-stirred 2D Bose gas}
\author{Vijay Pal Singh}
\affiliation{Zentrum f\"ur Optische Quantentechnologien, Universit\"at Hamburg, 22761 Hamburg, Germany}
\affiliation{Institut f\"ur Laserphysik, Universit\"at Hamburg, 22761 Hamburg, Germany}
\affiliation{The Hamburg Centre for Ultrafast Imaging, Luruper Chaussee 149, Hamburg 22761, Germany}
\author{Christof Weitenberg}
\affiliation{Institut f\"ur Laserphysik, Universit\"at Hamburg, 22761 Hamburg, Germany}
\author{Jean Dalibard}
\affiliation{Laboratoire Kastler Brossel, Coll\`ege de France, ENS-PSL Research University,
CNRS, UPMC-Sorbonne Universit\'es, 11 place Marcelin Berthelot, 75005 Paris, France}
\affiliation{The Hamburg Centre for Ultrafast Imaging, Luruper Chaussee 149, Hamburg 22761, Germany}
\author{Ludwig Mathey}
\affiliation{Zentrum f\"ur Optische Quantentechnologien, Universit\"at Hamburg, 22761 Hamburg, Germany}
\affiliation{Institut f\"ur Laserphysik, Universit\"at Hamburg, 22761 Hamburg, Germany}
\affiliation{The Hamburg Centre for Ultrafast Imaging, Luruper Chaussee 149, Hamburg 22761, Germany}
\date{\today}

\begin{abstract}
We investigate the superfluid behavior of a two-dimensional (2D) Bose gas of $^{87}$Rb atoms using classical field dynamics. In the experiment by R. Desbuquois \textit{et al.}, Nat. Phys. \textbf{8}, 645 (2012), a 2D quasicondensate in a trap is stirred by a blue-detuned laser beam along a circular path around the trap center. Here, we study this experiment from a theoretical perspective. The heating induced by stirring increases rapidly above a velocity $v_c$, which we define as the critical velocity. 
We identify the superfluid, the crossover, and the thermal regime by a finite, a sharply decreasing, and a vanishing critical velocity, respectively.
We demonstrate that the onset of heating occurs due to the creation of vortex-antivortex pairs.
A direct comparison of our numerical results to the experimental ones shows good agreement, if a systematic shift of the critical phase-space density is included. We relate this shift to the absence of thermal equilibrium between the condensate and the thermal wings, which were used in the experiment to extract the temperature. We expand on this observation by studying the full relaxation dynamics between the condensate and the thermal cloud.

%
\end{abstract}
\maketitle

\section{Introduction}

 Frictionless flow is one of the defining features of superfluidity \cite{Leggett}. 
For a moving obstacle with velocity $v$ in a superfluid, the frictionless nature of the superfluid near the obstacle breaks down when $v$ exceeds a certain critical velocity $v_c$. According to Landau's criterion this critical velocity is estimated as $v_c = \mathrm{min}_\bk [\epsilon(\bk)/\hbar k]$, where $\epsilon(\bk)$ is the excitation spectrum, $\hbar$ is the Planck constant, and $\bk$ is the wave vector, with $k=|\bk|$, see Refs. \cite{Leggett, PitaevskiiStringari, Pethick2008}.
An object moving with a velocity above $v_c$ dissipates energy via the creation of elementary excitations, for example, vortices or phonons. 
Superfluidity was first observed in liquid helium $4$ and helium $3$. 
Since then, superfluidity has been studied in quantum gas systems of bosons \cite{Ketterle1999,Dalibard2000,Atherton2007,Anderson2010, Dalibard2012}, fermions \cite{Zwierlein2011, Miller2007, Weimer2015}, as well as of Bose-Fermi mixtures \cite{Salomon2014}.

  The phenomenon of superfluidity is closely related to the Bose-Einstein condensation (BEC) of interacting gases. Interestingly, a uniform two-dimensional (2D) system cannot undergo the BEC transition because the formation of long-range order is precluded by thermal fluctuations \cite{Mermin1966, Hohenberg1967}. 
However, it forms a superfluid with quasi-long range order via the Berenzinskii-Kosterlitz-Thouless (BKT) mechanism \cite{Minnhagen1987}. 
The quasi-long range order of this state refers to the algebraic decay of the single-particle correlation function. The algebraic exponent of this correlation function increases smoothly with temperature. At the critical temperature, the superfluid density of the system undergoes a universal jump of $4/\lambda^2$, where $\lambda$ is the de Broglie wavelength. 
Experiments on 2D bosonic systems, such as a liquid helium film \cite{Bishop1978}, and  trapped Bose gases \cite{Dalibard2006, Clade2009, Tung2010, Plisson2011, Shin2013} have shown indications of the BKT transition.
Furthermore, a trapped 2D system can form a BEC due to the modified density of states \cite{Petrov2004, Hadzibabic2011} and leads to an interesting interplay of the two phase transitions \cite{Hadzibabic2015}.

 Quasi-long range order in 2D bosonic systems can be detected via interference and time-of-flight techniques \cite{Polkovnikov2006, Dalibard2006, Clade2009, Tung2010, Plisson2011, Shin2013, Shin2012, Singh2014}. 
However, as a direct method, superfluidity of ultracold atomic gases was probed using a local perturbation, in particular via laser stirring. 
For example, superfluidity of 3D BECs was tested via laser stirring in Refs. \cite{Ketterle1999, Weimer2015}. 
In the experiment \cite{Shin2012}, thermal relaxation of a perturbed 2D quasicondensate is studied.

  Ref. \cite{Dalibard2012} reported on stirring a trapped 2D Bose gas of $^{87}$Rb atoms with a blue-detuned laser, moving on a circular path around the trap center.
The circular motion ensures that the harmonically trapped 2D gas is probed at a fixed phase-space density.
By choosing different radii of the circular motion, the superfluid transition was explored.
In this paper, we provide a quantitative understanding of the experiment using a c-field simulation method. 
We demonstrate that a blue-detuned laser of intensity comparable to the mean-field energy causes dissipation due to the creation of vortex-antivortex pairs. 
This is in contrast to laser stirring with a red-detuned laser \cite{Weimer2015}, where dissipation occurs via phonons \cite{Singh2016}. 
Furthermore, we study the relaxation dynamics of the stirred gas following the stirring process, which shows a slow energy transport between the condensate and the thermal cloud. We identify the origin of this slow relaxation to be vortex recombination and diffusion. We show that this effect can explain quantitatively the shift of the measured critical phase-space density in the experiment.

  This paper is organized as follows. In Sec. \ref{sec_sim_method} we describe the simulation method that we use.
In Sec. \ref{sec_vc_sim} we determine the critical velocity $v_c$ of the stirred gas, based on which we identify the superfluid to thermal transition.
In Sec. \ref{sec_mechanism} we discuss the dissipation via vortex pairs. 
In Sec. \ref{sec_comp_expt} we compare the simulation results with the experiment.
In Sec. \ref{sec_nonequilibrium} we analyze the relaxation of the stirred gas, and in Sec. \ref{sec_conc} we conclude.

\section{Simulation method} \label{sec_sim_method}

\begin{figure*}[t]
\includegraphics[width=1.0\linewidth]{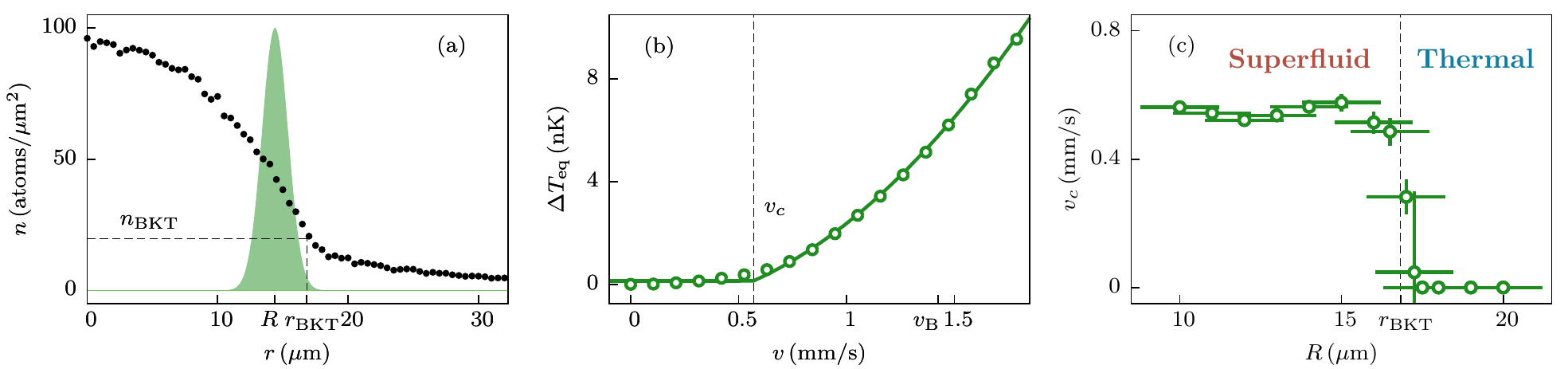}
\caption{\textbf{Determining the superfluid response and critical velocity.} 
In panel (a) we show the simulated density profile (dots) of the trapped 2D gas. We stir the gas with a repulsive Gaussian potential (green Gaussian beam of width $\sigma=1\, \mu \mathrm{m}$) on a circular path around the trap center at a stirring radius $R$. The phase-space density $\cD$ of the gas decreases with increasing $R$. The prediction for the BKT transition in a uniform gas \cite{Prokofev2001} yields the critical phase-space density $\cD_c =8.3$, which corresponds to the cloud density $n_\mathrm{BKT} = 19.8/\mu\mathrm{m}^{2}$ at $T= 85\, \mathrm{nK}$. Using local density approximation, the border of the superfluid region is expected to be at $r_{\mathrm{BKT}}=16.8\, \mu \mathrm{m}$.
In panel (b) we show the simulated heating $\Delta T_\meq$ (circles) as a function of stirring velocity $v = R \omega_m$, at $R=14.4\, \mu \mathrm{m}$. $\Delta T_\meq $ is determined from the equilibrium temperature $T_\meq$ of the stirred gas.
We determine a critical velocity $v_c$ of $0.58 \pm 0.2 \, \mathrm{mm/s}$ using the fitting function in Eq. \ref{fit}. We show $v_c$ and the fitted curve by the vertical dashed and the green continuous line, respectively. The Bogoliubov estimate of the phonon velocity at $R$ is $v_\mB = 1.4 \, \mathrm{mm/s}$. 
In panel (c) we show $v_c$ (circles) determined at various $R$.
The y errorbars show the standard deviation. The x errorbars denote the size of the stirring potential ($1/\sqrt{e}$ of diameter $2\sigma$).}
\label{fig_heating}
\end{figure*}

 We simulate the stirring dynamics of a weakly interacting 2D bosonic system using the c-field simulation method that we used for a 3D system in Ref. \cite{Singh2016}. We describe this method in the following. 
We start out with the Hamiltonian of the unperturbed system,
\begin{align} \label{eq_hamil}
\hat{H}_{0} &= \int \md \br \, \Big[ -  \hat{\psi}^\dagger({\bf r}) \frac{\hbar^2}{2m}  \nabla^2 \hat{\psi}({\bf r})  + V({\bf r}) \hat{\psi}^\dagger({\bf r})\hat{\psi}({\bf r}) \nonumber\\
&  \quad  + \frac{g}{2} \hat{\psi}^\dagger({\bf r})\hat{\psi}^\dagger({\bf r})\hat{\psi}({\bf r})\hat{\psi}({\bf r})\Big].
\end{align}
$\hat{\psi}$ and $\hat{\psi}^\dagger$ are the bosonic annihilation and creation operator, respectively. The 2D coupling parameter $g$ is given by $g= \tilde{g} \hbar^2/m$, where $\tilde{g}=\sqrt{8\pi} a_s/l_z$ is the dimensionless interaction, $m$ is the atomic mass, $a_s$ is the 3D $s$-wave scattering length, and $l_z= \sqrt{\hbar/(m \omega_z)}$ is the harmonic oscillator length of the confining potential $ m\omega_z^2 z^2/2$ in the $z$ direction. $\omega_z$ is the trap frequency along the $z$ direction.
$V( \br)$ describes the external potential, which is a harmonic trap, $V_h(\br) = m \omega_r^2r^2/2$. $\omega_r$ is the trap frequency in the radial direction and $r=(x^2+y^2)^{1/2}$ is the radial coordinate. 
We introduce a time-dependent term to describe laser stirring,
\begin{align} \label{eqn_hamiltonian_stir}
\hat{H}_{s}(t) = \int \md \br \, V({\bf r},t) \hat{n}({\bf r}),
\end{align} 
where $V({\bf r},t)$ is the time-dependent stirring potential and $\hat{n}({\bf r})$ is the density operator at the location $\br=(x,y)$.
The stirring potential is a Gaussian with a width $\sigma$ and a strength $V_0$,
\begin{equation} \label{eq_stir_potential}
V({\bf r}, t) = V_0 \exp \Bigl(- \frac{ \bigl( {\bf r} - {\bf r}_s(t) \bigr)^2} {2 \sigma^2} \Bigr),
\end{equation}
which is centered at ${\bf r}_s(t) = \bigl(x_s(t), y_s(t) \bigr)$. We move $(x_s, y_s)$ along a circular path as a function of time $t$. 

 We perform numerical simulations by mapping this system on a lattice system, which also introduces a short-range cutoff; see Appendix \ref{app_sim_heating}. This short-range cutoff is of the order of the healing length $\xi= \hbar/\sqrt{2mgn}$, with $n$ being the density.
We describe both the equations of motion and the initial state within a c-number representation, which corresponds to formally replacing the operators $\hat{\psi}$ by complex numbers $\psi$. Furthermore, we approximate the initial ensemble by a classical ensemble, within a grand-canonical ensemble of temperature $T$ and chemical potential $\mu$. We sample the initial states via a classical Metropolis algorithm.

 The simulation setup consists of a disc-shaped 2D circular condensate of $^{87}$Rb atoms. This choice of the 2D circular condensate is inspired by the experimental setup of Ref. \cite{Dalibard2012}. In the simulations we consider $N=38,162 - 93,267$ $^{87}$Rb atoms confined by the harmonic potential in both the radial and transverse directions. The trap frequencies are $\omega_r = 2\pi \times 25\, \mathrm{Hz}$ and  $\omega_z = 2\pi \times 1.5\, \mathrm{kHz}$. Here the scattering length is $a_s = 5.3\, \mathrm{nm}$, which yields $\tilde{g} =\sqrt{8\pi} a_s/l_z =0.093$. The temperature of the trapped gas is in the range $T = 63 - 85\, \mathrm{nK}$. 
The simulation parameters that we use, are in the typical range of the experimental parameters of Ref. \cite{Dalibard2012}. 
For simulations of a quasi- and a pure-2D trap geometry we use a lattice of $180 \times 180 \times 5$ and $200 \times 200$ sites, with the lattice discretization length $l = 0.5\, \mu \mathrm{m}$, respectively. We choose $l$ such that it is smaller than, or comparable to, the healing length $\xi$ and the de Broglie wavelength $\lambda =\sqrt{2\pi\hbar^2/mk_\mB T}$, see Ref.  \cite{Mora2003}. The trapped gas is in the pure-2D regime if $k_\mB T, \, \mu < \hbar \omega_z$. When $k_\mB T$ and $\mu$ are comparable to $\hbar \omega_z$, it is in the quasi-2D regime.

  After initializing the trapped system at temperature $T$, we switch on the stirring potential described by Eq. \ref{eq_stir_potential}. In the experiment \cite{Dalibard2012} the trapped gas is stirred with a blue-detuned laser beam moving on a circular path around the trap center. For the circular motion of stirring we choose $(x_s, y_s)= R \bigl(\cos(\omega_m t), \sin(\omega_m t) \bigr)$, where $R$ and $\omega_m$ are the stirring radius and frequency, respectively. For the stirring potential we use the strength $V_0= k_\mB \times 80\, \mathrm{nK}$ and the width $\sigma = 1\, \mu \mathrm{m}$, in accordance with the experiment.
The stirring sequence is the following: We linearly switch on the stirring potential over $5\, \mathrm{ms}$, let it stir the system for $200\, \mathrm{ms}$, and then switch it off over $5\, \mathrm{ms}$. This is again inspired by the experimental choices. 
We repeat this for various stirring velocities $v =R\omega_m$ by changing both $R$ and $\omega_m$. By choosing different $R$ we stir the different regimes of the trapped gas, the superfluid, the thermal, and the crossover regime.

 After stirring we calculate the total energy $E = \la H_{0}\ra$ using the unperturbed Hamiltonian in Eq. \ref{eq_hamil}, where we use $\psi$ instead of $\hat{\psi}$.
From this energy we determine the equilibrium temperature $T_\meq$ of the stirred gas. We infer this temperature by numerically inverting the temperature dependence of the equilibrium state, $E_\meq=E_\meq(T)$.
We elaborate on this in Appendix \ref{app_sim_heating}. 
From the temperature difference between the stirred and initial system, the heating $\Delta T_\meq = T_\meq- T$ is determined. 
We also calculate the local energy, as well as the vortex and anti-vortex distribution. 
We define the local energy as $E_i = -\frac{J}{2} \sum_j (\psi_i^\ast \psi_j + \psi_j^\ast \psi_i) + \frac{U}{2}n_i^2 +V_i n_i$, where $j$ refers to the nearest neighbor sites. $\psi_i$, $n_i= |\psi_i|^2$, and $V_i$ are the complex-valued field, the density, and the trap potential  at site $i$, respectively. $J$ and $U$ are the Bose-Hubbard parameters, see Appendix \ref{app_sim_heating}. 
For the vortex distribution, we calculate the phase winding around the lattice plaquette of size $l\times l$, using $\sum_{\Box} \delta \phi(x,y) = \delta_x\phi(x,y) + \delta_y\phi(x+l,y)+\delta_x\phi(x+l,y+l)+\delta_y\phi(x,y+l)$, where the phase differences between sites 
is taken to be $\delta_{x/y} \phi(x,y)  \in (-\pi, \pi]$. 
$\phi$ is the phase field of $\psi$.
We identify a vortex and an antivortex by a phase winding of $2\pi$ and $-2\pi$, respectively. 
By counting all vortices and antivortices we determine the total number of vortices. We restrict this counting to the the superfluid region of the gas as we describe below.

\section{Superfluid response} \label{sec_vc_sim}

\begin{figure*}[t]
\includegraphics[width=0.95\linewidth]{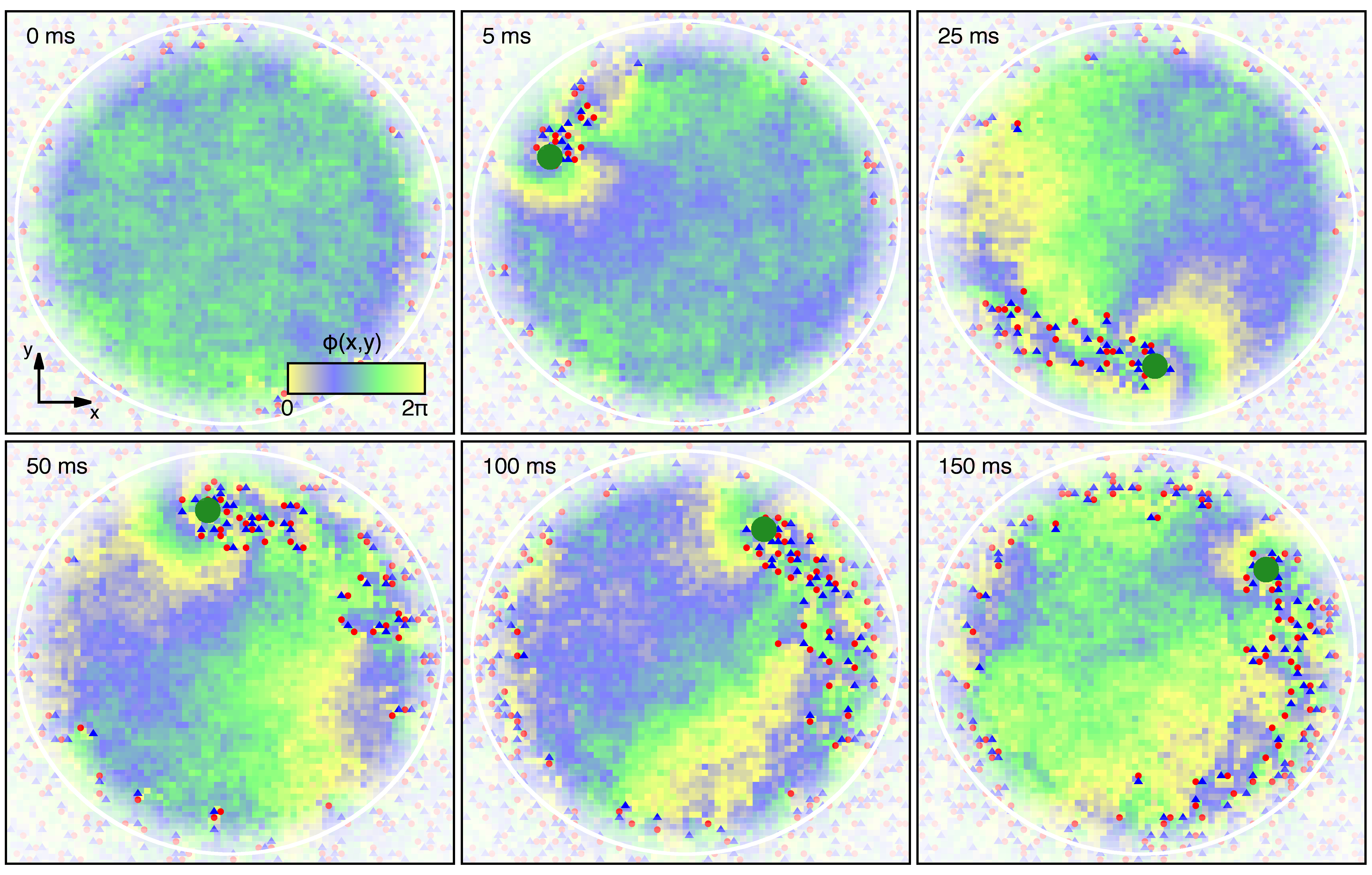}
\caption{\textbf{Dissipation due to vortex-antivortex pairs.} 
The phase evolution of a single realization of the trapped gas stirred with a repulsive Gaussian potential along a circular path at the times $t=5,\, 25,\, 50,\, 100,\, 150\,  \mathrm{ms}$. The stirring potential (green disc of diameter $2\sigma$) moves counterclockwise in a circle at $R=12\, \mu \mathrm{m}$ with a velocity $v> v_c$.
This creates strong phase gradients, which result in the creation of vortex-antivortex pairs. We identify a vortex (red circle) and an antivortex (blue triangle) by a phase winding of $2\pi$ and $-2\pi$ around the lattice plaquette, respectively. The white line indicates the superfluid-thermal boundary circle, based on $r_\mBKT$. In the thermal region (white shaded region outside of $r_\mBKT$) the phase fluctuates strongly. The field of view in each figure is of size $35\, \mu \mm \times 35\, \mu \mm$.
}
\label{fig_dissipation}
\end{figure*}

\begin{figure*}
\includegraphics[width=1.0\linewidth]{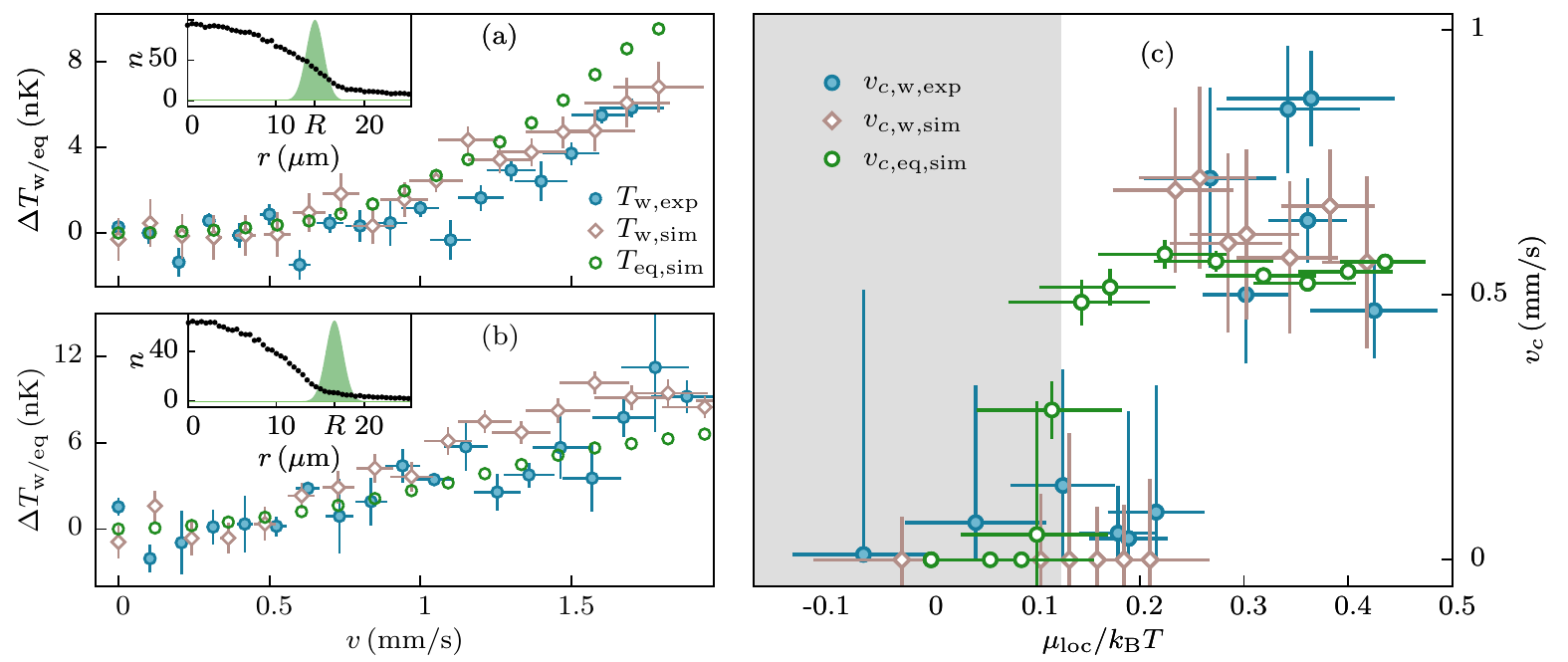}
\caption{\textbf{Comparison of simulation and experimental results.} 
In panel (a) we stir the quasicondensate region at $R=14.4\, \mu \mathrm{m}$ (inset), whereas in panel (b) we stir the thermal region at $R=16.6\, \mu \mathrm{m}$ (inset). 
We compare the measured heating $\Delta T_\mw$ (solid circles) with the simulated $\Delta T_\mw$ (diamonds) and $\Delta T_\meq$ (open circles) at varying stirring velocity $v$. $\Delta T_\mw$ and $\Delta T_\meq$ are determined from the wing temperature $T_\mw$ and equilibrium temperature $T_\meq$, respectively.
The y errorbars are the standard deviation. The x errorbars indicate the spread of velocities along the size of the stirring potential. 
In panel (c) we compare the measured $v_{c, \mw}$ (solid circles) with the simulated $v_{c, \mw}$ (diamonds) and $v_{c, \meq}$ (open circles) across the superfluid-thermal transition.
According to the BKT prediction in a uniform gas \cite{Prokofev2001} this transition occurs at $(\mu_{\mathrm{loc}} /k_\mB T)_c \approx 0.13$. The thermal state by this prediction is indicated by the gray shaded area.
The x errorbars denote the region of $\mu_{\mathrm{loc}} /k_\mB T$, which is probed by the stirring potential due to its size. The y errorbars show the standard deviation. The experimental data is from Ref. \cite{Dalibard2012}. }
\label{fig_heating_comp_SF}
\end{figure*}

 To study the superfluid behavior we stir a 2D quasicondensate with a repulsive Gaussian potential.
We prepare a trapped 2D quasicondensate of $N= 93,267$ $^{87}$Rb atoms at temperature $T= 85\, \mathrm{nK}$.
We show the simulated density profile of the trapped gas in Fig. \ref{fig_heating}(a). 
We stir the gas with a circularly moving, repulsive stirring potential at stirring radius $R=14.4\, \mu \mathrm{m}$. As mentioned in Sec. \ref{sec_sim_method}, we use the strength $V_0= k_\mB \times 80\, \mathrm{nK}$ and the width $\sigma = 1\, \mu \mathrm{m}$ for the stirring potential. This strength $V_0$ is well above the local mean-field energy $\mu_{\mathrm{loc}} \approx k_\mB \times 21\, \mathrm{nK}$ at the stirring location. 
After stirring we determine the induced heating $\Delta T_\meq = T_{\meq} -T$ from the equilibrium temperature $T_{\meq}$ of the stirred gas, see Sec. \ref{sec_sim_method} for details. By varying the stirring frequency $\omega_m$ we determine $\Delta T_{\meq}$ as a function of stirring velocity $v=R\omega_m$. We show $\Delta T_\meq $ determined for various $v$ in Fig. \ref{fig_heating}(b). 
The induced heating is almost negligible at low $v$, its onset occurs at a velocity $v_c$, and for $v>v_c$ it increases rapidly.
We quantify the onset of heating using a fitting function, 
\begin{equation} \label{fit}
(\Delta T)_{\mathrm{fit}} = A \cdot \max[0, (v^2-v_c^2)] + B,
\end{equation}  
which is discussed in Ref. \cite{Pitaevskii2004}, with the free parameters $A$, $B$, and $v_c$. For the simulated heating shown in Fig. \ref{fig_heating}(b), this function gives a critical velocity of $v_c = 0.58 \pm 0.2 \, \mathrm{mm/s}$. We compare this critical velocity to the Bogoliubov estimate of the phonon velocity $v_\mB = 1.4 \, \mathrm{mm/s}$ at the stirrer location. The Bogoliubov velocity is determined by $v_\mB = \hbar \sqrt{\tilde{g}n}/m$. 
The observed critical velocity is $v_c \approx 0.4\, v_\mB$. This is notably different from the case of an attractive stirring potential, where $v_c \approx v_\mB$ \cite{Singh2016}. We explain this reduction of $v_c$ for a repulsive stirring potential in Sec. \ref{sec_mechanism}.

  By choosing different radii $R$ we explore the various regimes of the trapped gas. 
We use the same strength $V_0$ and the same width $\sigma$ as above.
For each $R$, we first determine the induced heating $\Delta T_{\meq}$ as a function of $v$, and then by using the fitting function given in Eq. \ref{fit} we determine $v_c$. We show $v_c$ determined at various $R$ in Fig. \ref{fig_heating}(c).
The stirring radii are in the range $R=10-20\, \mu \mm$. For $R=10-16\, \mu \mm$, there is no significant change of $v_c$. 
As $R$ reaches the crossover regime, $v_c$ is reduced sharply and for $R$ above the crossover regime, $v_c$ is zero. According to the BKT prediction in a uniform system \cite{Prokofev2001} with $\tilde{g}=0.093$ combined with local density approximation, the crossover regime should occur at $r_\mathrm{BKT}=16.8\, \mu \mm$. 
This prediction is in good agreement with the crossover regime identified by the simulated $v_c$. 
Thus, we clearly identify the superfluid, the crossover, and the thermal regimes by the finite, the sharply decreasing, and the zero critical velocities $v_c$, respectively.  
We note that in the crossover region the decrease of $v_c$ is as sharp as the size of the stirrer allows.
Furthermore, we note that the observed almost constant $v_c$ for $R < r_\mathrm{BKT}$ can be due to the accelerated circular motion and the large strength of the stirring potential \cite{Singh2016}.


\section{Dissipation mechanism} \label{sec_mechanism}

 The observed critical velocities are in the range $v_c=0.3-0.5\, v_\mB$. To understand what leads to this reduction of the critical velocity with regard to the phonon velocity, we investigate the time evolution of the phase field $\phi (\br, t)$ of a single realization of the thermal ensemble. We obtain this phase field from the complex field $\psi (\br)$ via the phase-density representation $\psi = \sqrt{n} \exp(i \phi)$. 
In Fig. \ref{fig_dissipation} we show the phase evolution of the trapped 2D quasicondensate stirred at $R=12\, \mu \mathrm{m}$. We use the velocity $v \approx 0.8\, v_\mB$, which is above the steep onset of dissipation related to the breakdown of superfluidity.   
The phase evolution of the unperturbed gas shows rather weak phase gradients. 
As stirring is switched on, the phase field around the stirring potential starts to fluctuate. These fluctuations develop into strong phase gradients, which result in the creation of vortex-antivortex pairs.
This can be confirmed by calculating the phase winding around each plaquette of our numerical grid, as described in Sec. \ref{sec_sim_method}. 
We show the calculated phase winding in Fig. \ref{fig_dissipation}, where vortices and antivortices are shown as circles and triangles, respectively. 
This indeed shows the creation of vortex-antivortex pairs during stirring.
We recall that the stirring strength $V_0= k_\mB \times 80\, \mathrm{nK}$ is much larger than the mean-field energy $\mu_{\mathrm{loc}} \approx k_\mB \times 30\, \mathrm{nK}$ at the stirring location, which results in a strong reduction of the density at the stirrer location. This density reduction serves as a nucleation site for the creation of
vortex pairs. 
We note that this mechanism of vortex-pair-induced dissipation is suppressed for an attractive stirring potential, as shown in Ref. \cite{Singh2016}. This scenario of dissipation induced by vortex pair creation is consistent with a recent experiment \cite{Shin2015}.

\begin{figure*}
\includegraphics[width=0.95\linewidth]{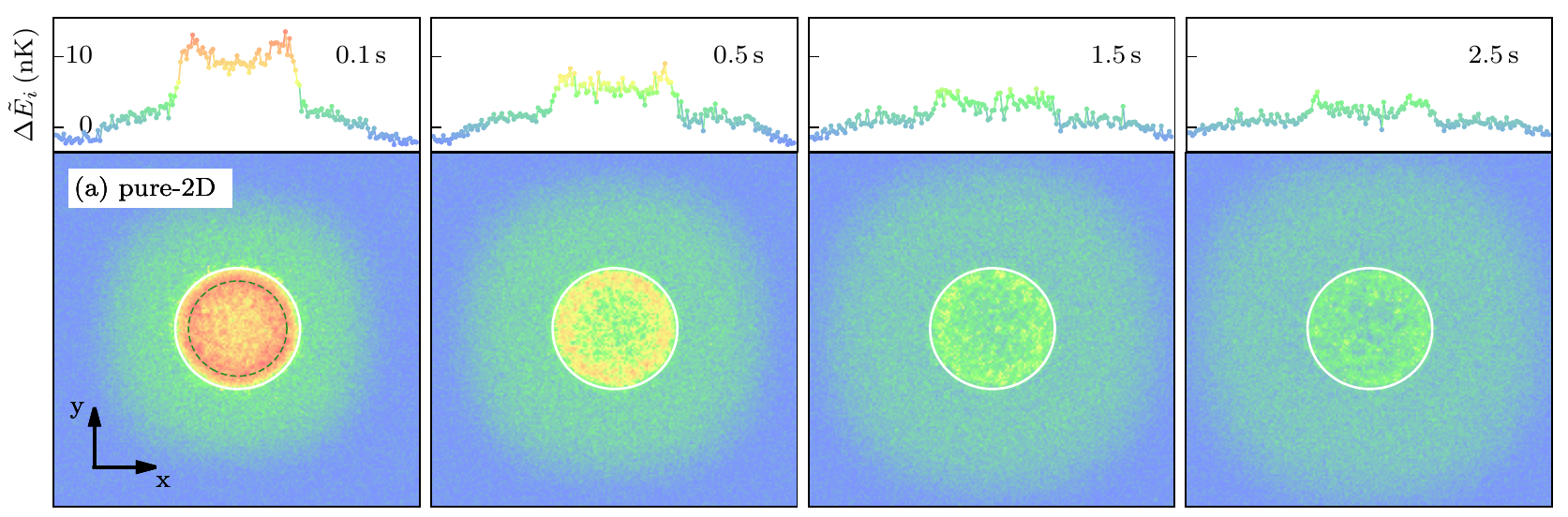} \\
\includegraphics[width=0.95\linewidth]{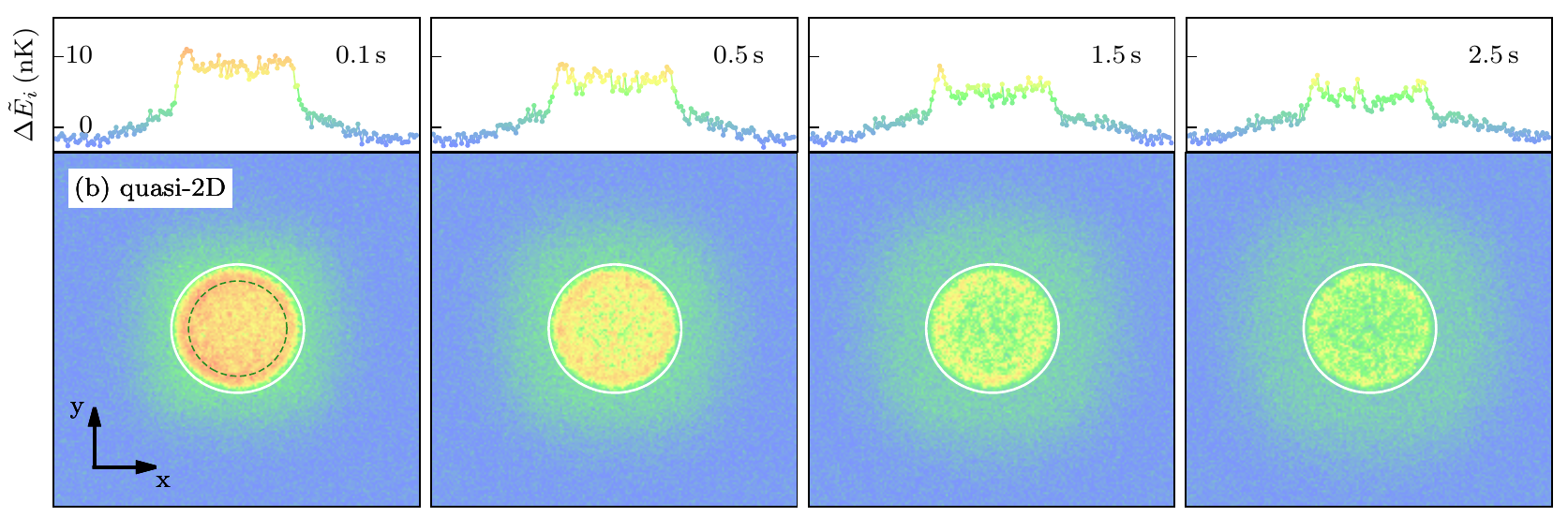} 
\caption{\textbf{Relaxation dynamics.}
Energy flow of the stirred trapped gas for the relaxation times $t_{\mathrm{relax}} = 0.1,\, 0.5, \, 1.5, \, 2.5\, \ms$. The gas was stirred with a repulsive Gaussian potential at $R=12\, \mu \mm$ (circle indicated by the green dashed line).
We show the evolution of the excess energy $\Delta \tilde{E}_i= \bigl(E_i(t) -E_i^\meq\bigr)/n_\mmax$ for the pure-2D gas in the upper panels (a) and for the quasi-2D gas in the lower panels (b). $E_i$ is the local energy of the stirred gas at time $t$ and $E_i^\meq$ is its final equilibrium local energy. $n_\mmax$ is the maximum density of the system. The top panels of (a) and (b) are the one-dimensional cut through the trap center. 
The superfluid-thermal boundary circle is indicated by the white line. $\Delta \tilde{E}_i$ in the far wings of the cloud is negative as $E_i^\meq$ is larger than $E_i$.
The field of view is $90\, \mu \mm \times 90\, \mu \mm$.
}
\label{fig_energyevolution}
\end{figure*}

\section{Comparison to experiment} \label{sec_comp_expt}

  We now compare the results of our simulation with the experiment \cite{Dalibard2012}. 
We first show the comparison between the experiment and simulation for the heating as a function of $v$.
In the superfluid regime, we stir the quasicondensate at the radius $R=14.4\, \mu \mathrm{m}$. The simulated density profile is shown in the inset of Fig. \ref{fig_heating_comp_SF}(a).
After stirring we let the stirred gas relax for $100\, \mathrm{ms}$ of relaxation time and then determine the induced heating from the temperature of the wings of the cloud. We fit these wings  to the Hartree-Fock prediction, 
\begin{equation}\label{eq_HF_fit}
n(\br) = - \frac{m k_\mB T}{2\pi \hbar^2} \ln \Bigl[1 - \exp \Bigl( \frac{\mu_0 - V_h(\br)- 2 g n(\br) }{k_\mB T} \Bigr)  \Bigr],
\end{equation}
with the fitting parameters $T$ and $\mu_0$. This method is adopted according to experiment, in which the temperature of the stirred gas is determined in the same way, following a relaxation of $100\, \mathrm{ms}$ as well.
We denote this heating determined from the wing temperature $T_\mw$ by $\Delta T_\mw = T_\mw -T$, with $T$ being the initial temperature.
We show the simulated $\Delta T_\mw$ and their comparison with the experimental ones for various $v$ in Fig. \ref{fig_heating_comp_SF}(a).
The measured and simulated heating are found to be in good agreement if we base the comparison on $\Delta T_\mw$. 
We also compare the measured $\Delta T_\mw$ with the simulated $\Delta T_\meq$ determined from the equilibrium temperature $T_\meq$ of the stirred gas. We show $\Delta T_\meq$ as open circles in Fig. \ref{fig_heating_comp_SF}(a).
They show agreement at low and intermediate velocities $v$, whereas they differ at large $v$.  
This noticeable difference at large $v$ is due to the absence of global thermal equilibrium of the stirred gas.
As explained in Sec. \ref{sec_nonequilibrium}, the stirred gas relaxes by transporting the excess energy between the superfluid in the central part and the thermal cloud in the periphery, which is a slow process. The absence of global thermal equilibrium leads to a smaller wing temperature than the equilibrium temperature.

   The results shown in Fig. \ref{fig_heating_comp_SF}(a) indicate that the onset of heating occurs at a velocity $v_c$, and for $v > v_c$ heating increases rapidly. Both in experiment and simulation $v_c$ is determined using the fitting function in Eq. \ref{fit}. 
In Fig. \ref{fig_heating_comp_SF}(b) we show the comparison between the experiment and simulation for stirring  the thermal region of the trapped gas of $N=38,162$ atoms. The simulated density profile of the gas is given in the inset of Fig. \ref{fig_heating_comp_SF}(b). Both the measured and simulated heating  $\Delta T_\mw$ are in good agreement. The simulated $\Delta T_\meq$ determined from the equilibrium temperature of the system are below the measured  $\Delta T_\mw$ at large $v$. 
As we will explain in Sec. \ref{sec_nonequilibrium}, this is again due to the absence of global thermal equilibrium. As the stirred thermal cloud has more excess energy than the condensate, the wing temperature is larger than the equilibrium temperature. The results shown in Fig. \ref{fig_heating_comp_SF}(b) indicate that heating occurs at all $v$, which results in a zero $v_c$.

 Next, we show in Fig.  \ref{fig_heating_comp_SF}(c) the comparison between the experiment and simulation for $v_c$ that are determined by stirring the superfluid, the crossover, and the thermal regime. In the experiment \cite{Dalibard2012} $v_c$ is measured for different configurations of the total number of atoms $N$, the temperatures $T$, and the stirring radii $R$. We compare the measured $v_c$ with the simulated $v_c$ determined by stirring the 2D gas in Sec. \ref{sec_vc_sim}. We show both the measured and simulated $v_c$ as a function of the dimensionless parameter $\mu_{\mathrm{loc}}(r)/k_\mB T$. 
The parameter $\mu_{\mathrm{loc}}/k_\mB T$ characterizes the degree of degeneracy of the cloud and is the relavant parameter in the sense that the thermodynamic properties of the gas depend only on the ratio $\mu/k_\mB T$ \cite{Yefsah2011, Prokofev2001, Hung2011}. 
We refer to $v_c$ determined from the wing temperature $T_\mw$ and from the equilibrium temperature $T_\meq$ as $v_{c, \mw}$ and $v_{c, \meq}$, respectively. 
Both the measured and simulated $v_{c, \mw}$ show good agreement. The measured $v_{c, \mw}$ and the simulated $v_{c, \meq}$ agree in the superfluid and thermal regime, while they differ in the crossover regime. 
For the measured and simulated $v_{c, \mw}$, the crossover regime occurs at $\mu_{\mathrm{loc}}/k_\mB T  \approx 0.24$ and $0.22$, respectively. However, for the simulated $v_{c, \meq}$, it occurs at $\mu_{\mathrm{loc}}/k_\mB T  \approx 0.11$. 
The theoretical prediction for the BKT transition in a uniform gas \cite{Prokofev2001} with $\tilde{g}= 0.093$ occurs at $(\mu/k_\mB T)_c \approx 0.13$. This prediction is comparable to the simulated crossover regime identified by $v_{c, \meq}$, whereas its comparison with the crossover regimes identified by the measured and simulated $v_{c, \mw}$ shows a shift. 
This shift was observed in Ref. \cite{Dalibard2012}, but could not be explained. 
We conclude that the experiments of Ref. \cite{Dalibard2012} can be reproduced quantitatively if the wing temperature is used, rather than $T_\meq$. This suggests that the system has not relaxed to thermal equilibrium after the waiting time of $100\, \mathrm{ms}$. We confirm and elaborate on this point and the underlying mechanism in the following sections.

\begin{figure*}
\includegraphics[width=0.95\linewidth]{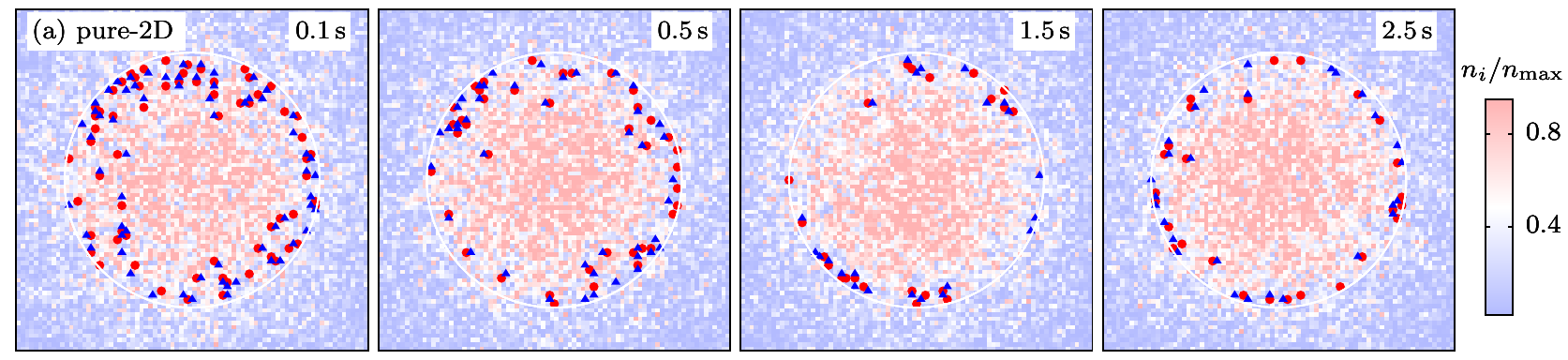}\\
\includegraphics[width=0.95\linewidth]{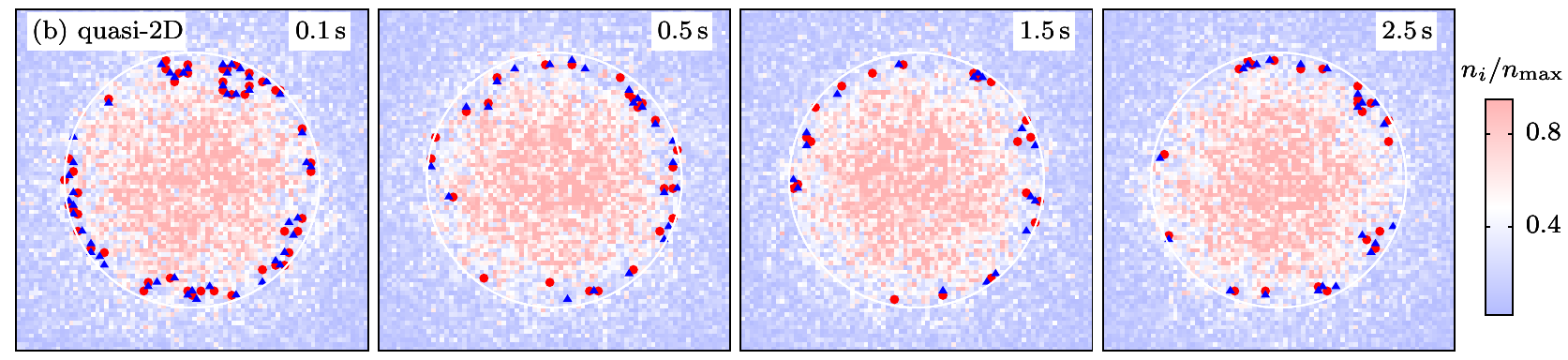}
\caption{\textbf{Relaxation of the density and vortices.} 
We show the evolution of the density and vortices of a single realization of the stirred gas for $t_{\mathrm{relax}} = 0.1,\, 0.5, \, 1.5, \, 2.5\, \ms$. The upper panels (a) and lower panels (b) correspond to the pure- and quasi-2D gas, respectively. For the quasi-2D gas we show the column density and the vortex distribution of the central ($z=0$) $xy$ plane. The maximum density $n_\mmax$ is $71/\mu \mm^2$ for pure-2D gas and the maximum column density $n_\mmax$ is $95/\mu \mm^2$ for quasi-2D gas.
We show vortices (red circles) and antivortices (blue triangles) in the superfluid region at the detection radius $R_\mdet = 14.5\, \mu \mm$ (circle indicated by the white line), which is below $r_\mBKT$ \cite{remark1}. 
The field of view is $40\, \mu \mm \times 40\, \mu \mm$.
 }
\label{fig_vortexdynamics}
\end{figure*}

\section{Relaxation dynamics} \label{sec_nonequilibrium}
 We now investigate the relaxation of the system, following the stirring process in the superfluid regime. This includes a discussion of the influence of the confinement of the system in the $z$ direction. For strong confinement, the system approaches a purely 2D limit, while it is quasi-2D for intermediate confinement.

\subsection{Energy-flow dynamics} \label{sec_energyflow}

 We first analyze the energy-flow dynamics of a stirred trapped gas in the purely 2D limit, and then compare this dynamics with a quasi-2D gas.
For a pure-2D trapped gas, we consider a gas of $N= 64,079$ $^{87}$Rb atoms, which is strongly confined in the transverse direction by the harmonic potential. The temperature $T= 85\, \mathrm{nK}$ is smaller than the transverse trap energy $\hbar \omega_z/k_\mB = 144\, \mathrm{nK}$, so that the gas is in the ground state in this direction. As the width of the condensate in the $z$ direction is smaller than the lattice discretization length $l$, we simulate this system using a single $xy$-layer of lattice only, see Sec. \ref{sec_sim_method}. 
We stir the gas at $R=12\, \mu \mm$ for $200 \, \mathrm{ms}$ at a velocity $v>v_c$. After that we switch off the stirring potential and let the gas relax.
We calculate the local energy $E_i$ of the stirred gas and its final equilibrated local energy $E_i^\meq$, as described in Sec. \ref{sec_sim_method}.
We show the evolution of the excess energy $\Delta \tilde{E}_i = \bigl(E_i(t) - E_i^\meq \bigr)/n_\mmax$ for various relaxation times $t_{\mathrm{relax}}$ in Fig. \ref{fig_energyevolution}(a). $n_\mmax$ is the maximum density of the system. 
The evolution of $\Delta \tilde{E}_i$ after $t_{\mathrm{relax}} = 0.1\, \ms$ shows that most of the stirring-induced energy resides within the superfluid region.
The system then relaxes by transporting this excess energy to the thermal cloud. This process occurs slowly and the system achieves fully equilibration only after about $2.5\, \ms$ relaxation time, remarkably.

 We now study this energy-flow dynamics for the case of a quasi-2D gas.
We consider the quasicondensate that we use in Sec. \ref{sec_vc_sim}. The initial temperature of the gas and the harmonic potential in the transverse direction are equal and half of those in the pure-2D case, respectively. The resulting system is a quasi-2D gas.
We simulate this system using five $xy$-layers of lattice in the $z$ direction.
We stir the gas using the same stirring parameters as for the pure-2D case.
We show the evolution of the excess energy $\Delta \tilde{E}_i$ of the stirred gas for various $t_{\mathrm{relax}}$ in Fig. \ref{fig_energyevolution}(b). In this case, $E_i$, $E_i^\meq$ and $n_\mmax$ are the column (i.e. integrated along the $z$ axis) quantities. 
The evolution of $\Delta \tilde{E}_i$ after $t_{\mathrm{relax}} = 0.1\, \ms$ is similar to the pure-2D gas. 
Again, the system relaxes by transporting the excess energy to the thermal cloud.
The equilibration process is slightly slower than for the pure-2D gas but again of the same order of $2.5\, \ms$.
We note that the measured relaxation times in Ref. \cite{Shin2012} are indeed of the order of the relaxation times that we find here.

  In the experiment \cite{Dalibard2012} the waiting time after stirring and before measurement is $0.1\, \ms$, which is shorter than the relaxation times that we observe here.
This indicates that in the experiment thermal equilibrium between the superfluid and thermal cloud is not fully established, which influences the measured heating. 
It leads to respectively lower and higher measured heating for stirring the superfluid and thermal parts of the cloud.

\subsection{Vortex dynamics} \label{sec_vortex}

  To understand what causes this slow relaxation for the system, we now examine the evolution of the density and vortices of the system. We calculate the local density as $n_i= |\psi_i|^2$ and vortices as described in Sec. \ref{sec_sim_method}.  
We show the density and vortex evolution of a single realization of the stirred pure- and quasi-2D gas after various $t_{\mathrm{relax}}$ in Fig. \ref{fig_vortexdynamics}.
For both systems, the density relaxation is hard to recognize, whereas the vortex evolution clearly exhibits decay of vortices. 
Thus, the system relaxes via decay of the induced vortices.
Vortices can decay via both annihilation of a vortex with an antivortex, and drifting out to the thermal region of the cloud.
For the pure-2D gas the number of vortices after $t_{\mathrm{relax}} = 0.1\, \ms$  is larger than the quasi-2D gas. 
Ref. \cite{Anderson2010, Rooney2011} reported that vortex annihilation in a pure-2D system is strongly suppressed as compared to a quasi-2D system because the vortex lines are impermeable to tilting \cite{Haljan2001} and bending \cite{Bretin2003}. 
So, the suppression of vortex annihilation can be a reason for the long-lived vortices for the pure-2D gas.

\begin{figure}
\includegraphics[width=1.\linewidth]{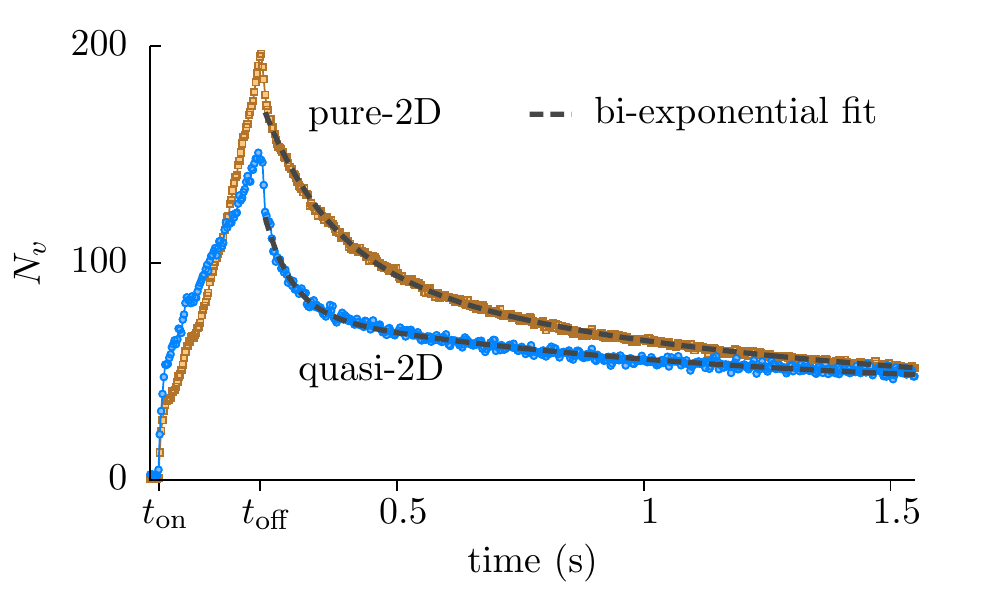}
\caption{\textbf{Vortex relaxation.} 
 We show the vortex number $N_v$ as a function of time $t$ for the pure- and quasi-2D gas. 
We fit the vortex decay with the bi-exponential function in Eq. \ref{eq_fit_decay} and determine the decay times $\tau_{d_1},\, \tau_{d_2}$ (see text). The fitted curves are shown as the dashed lines.
}
\label{fig_vortexdecay}
\end{figure}

  To make a quantitative comparison for the vortex relaxation between the two system, we count the total number of vortices within the superfluid region of the cloud at a detection radius $R_\mdet = 14.5\, \mu \mm$, and average it over $200$ realizations.
We show the averaged vortex number $N_v$ as a function of time $t$ for both systems in Fig. \ref{fig_vortexdecay}.    
As stirring is switched on at time $t_\mathrm{on}$, $N_v$ starts to increase approximately linearly. It reaches its maximum $N_\mmax$ at $t_\mathrm{off}$. 
After the stirring is switched off, it decays approximately exponentially.
For both systems the nature of vortex growth and decay are the same, but the rates with which they grow and decay are different. 
For the pure-2D gas the growth and decay rates are larger and smaller than those for the quasi-2D gas, respectively.
The enhanced growth and the suppressed decay rate for the pure-2D gas can be due to the suppression of vortex annihilation, as mentioned above, and a slow vortex drift. 
We quantify the vortex decay rate using the function, 
\begin{align} \label{eq_fit_decay}
f(t)=  N_1 e^{-t/\tau_{d_1}} + N_2 e^{-t/\tau_{d_2}} +N_0,
\end{align}
with the free parameters $N_1, \, N_2, \, N_0, \, \tau_{d_1},\, \tau_{d_2}$.  
From the fit, we determine $\tau_{d_1},\, \tau_{d_2} \approx 0.13,\, 0.87\, \ms$ and  $0.05,\, 0.87\, \ms$ for pure- and quasi-2D gas, respectively. 
These decay times are similar to those determined from the mean excess energy in Appendix \ref{app_energy_flow}.
The fast decay $\tau_{d_1}$ and the slow decay $\tau_{d_2}$ are essentially connected to the vortex annihilation and drift lifetime, respectively. 
For the pure-2D gas $\tau_{d_1}$ and $\tau_{d_2}$ are larger than and equal to those for the quasi-2D gas, respectively.  

  We show in Table \ref{tab:Rdet} $N_\mmax$ and the extracted $\tau_{d_1},\, \tau_{d_2}$ at varying $R_\mdet$, for both systems. $\tau_{d_1}$ and $\tau_{d_2}$ increase weakly as $R_\mdet$ is increased. However, the following conclusions are essentially independent of the choice of $R_\mdet$. Overall, $N_\mmax$ and $\tau_{d_1}$ are larger for pure-2D gas than those for quasi-2D gas, respectively, while $\tau_{d_2}$ are similar for both systems.
We compare the simulated $\tau_{d_1},\, \tau_{d_2}$ of quasi-2D gas to the waiting time  of $0.1\, \ms$ in the experiment \cite{Dalibard2012}. This time is twice as large as the fast decay $\tau_{d_1}$, whereas it is smaller than the slow decay $\tau_{d_2}$. 
This suggests that most vortex recombination processes have occurred at the time of the measurement. However, the vortex drift to the thermal cloud has not occurred, and the system is in a metastable state, not in the equilibrated state. This is the mechanism that is responsible for the difference between the wing temperature and the equilibrium temperature.


\begin{table}
\caption{$N_\mmax$ and the extracted $\tau_{d_1},\, \tau_{d_2}$ for different $R_\mdet$. } 
\begin{tabular}{c c c c c c c}
\hline\hline 
\multicolumn{3}{c}{ \qquad Pure-2D} &  \multicolumn{3}{c}{\qquad Quasi-2D}\\
\cline{2-7}  
$R_\mdet \, (\mu \mm)$ \quad & $N_{\mmax}$ \quad &  $\tau_{d_1}\, (\ms)$ \quad & $\tau_{d_2}\, (\ms)$ \quad &
$N_{\mmax}$ \quad &  $\tau_{d_1}\, (\ms)$ \quad  & $\tau_{d_2}\, (\ms)$ \\[0.5ex] 
\hline 
14.0 & 172 & 0.129 & 0.809 & 127 & 0.050 &  0.806  \\ 
14.5 & 196 & 0.135 & 0.866 & 151 & 0.055 &  0.869 \\
15.0 & 224 & 0.142 & 0.920 & 182 & 0.059 &  0.933 \\
15.5 & 251 & 0.147 & 0.959 & 214 & 0.066 &  0.989 \\  [1ex] 
\hline\hline 
\end{tabular}
\label{tab:Rdet}
\end{table}

  We note that in Ref. \cite{Fedichev1999} an estimate for the time of a vortex line drifting to the thermal cloud was given. While this estimate was for a three dimensional system, we find that the analytical estimate of Ref. \cite{Fedichev1999} gives a timescale that is consistent with our simulation.
We also note that the vortex lifetime is suppressed at high tempearures \cite{Ketterle2002, Jackson2009, Moon2015}.

\section{Conclusions} \label{sec_conc}

 We have studied the superfluid to thermal transition of a trapped 2D Bose gas of $^{87}$Rb atoms by stirring it with a repulsive stirring potential on a circular path around the trap center. The superfluid transition was probed by choosing different radii of the circular motion. 
We have identified the superfluid, the crossover, and the thermal regime by the finite, the sharply decreasing, and the zero critical velocity, respectively.    
The superfluid region of the gas yields critical velocities that are in the range $v_c= 0.3-0.5\, v_\mB$, where $v_\mB$ is the phonon velocity. 
We have demonstrated that the onset of dissipation is due to the creation of vortex-antivortex pairs.
The comparison of the simulation with the experiment shows good agreement if the temperature measurement of the experiment is imitated in the simulation, i.e. by extracting the wing temperature. However, we confirm the systematic shift that was observed in experiment, if thermal equilibrium is assumed.
 We have demonstrated that the absence of thermal equilibrium after the waiting time that was used in experiment is due to a remarkably slow relaxation mechanism: The energy transport across the superfluid to thermal interface occurs only on timescales of seconds. This slow transport mechanism is due to the slow drift of vortices out of the superfluid into the thermal wings of the system. We emphasize that this mechanism is relevant for many on-going experiments in the field of ultracold atoms, and their temperature measurements. Furthermore, this  effect of suppressed transport across critical interfaces is in itself intriguing, and could be studied in a future cold atom experiment with clarity.

\section*{acknowledgements}
 We acknowledge support from the Deutsche Forschungsgemeinschaft through Grants No. MA 5900/1-1 and No. SFB 925, the Hamburg Centre for Ultrafast Imaging, and from the Landesexzellenzinitiative Hamburg, supported by the Joachim Herz Stiftung.
JD thanks J\'er\^ome Beugnon for many fruitful discussions and acknowledges support from ERC (Synergy grant UQUAM).

\appendix

\section{Simulated heating}\label{app_sim_heating}

 In this section we show how we determine the equilibrium temperature $T_\meq$ of a stirred gas using the c-field method described in Sec. \ref{sec_sim_method}. We discretize the continuum Hamiltonian in Eq. \ref{eq_hamil} by the Bose-Hubbard Hamiltonian \cite{Jaksch1998} on a 2D square lattice, 
\begin{equation} \label{eq_hubbard}
H_0 = -J \sum_{\la i j \ra} (\psi_i^\ast \psi_j + \psi_j^\ast \psi_i) + \frac{U}{2} \sum_i n_i^2
 + \sum_i V_i n_i.
\end{equation}
$\psi_i$ and $n_i= |\psi_i|^2$ are the complex-valued field and the density at site $i$, respectively. $\la ij \ra$ indicates nearest-neighbor bonds. For a lattice discretization length $l$, the Bose-Hubbard parameters are related to the continuum parameters via $J= \hbar^2/(2ml^2)$ and $U=gl^{-2}$. The 2D coupling parameter $g$ is given by $g= \tilde{g} \hbar^2/m$, where $\tilde{g}= \sqrt{8 \pi} a_s/l_z$ is the dimensionless interaction, $m$ is the atomic mass,  $a_s$ is the 3D s-wave scattering length, and $l_z = \sqrt{\hbar/(m \omega_z)}$ is the harmonic oscillator length of the confining potential $m \omega_z^2z^2/2$ in the $z$ direction. $\omega_z$ is the trap frequency in the $z$ direction. The harmonic trapping potential is $V_i = m \omega_r^2 r^2/2$. $\omega_r$ is the trap frequency in the radial direction and $r= (x^2+y^2)^{1/2}$ is the radial coordinate.       

\begin{figure}
\includegraphics[width=0.95\linewidth]{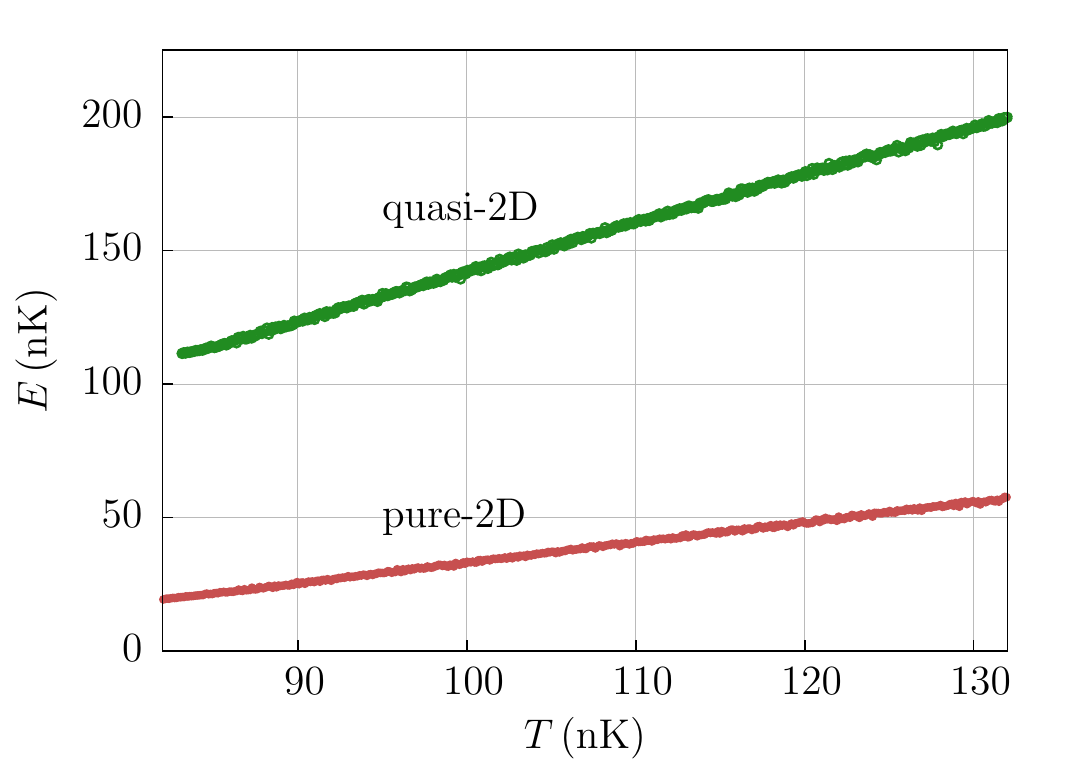}
\caption{Temperature dependence of the energy per atom for the pure- and quasi-2D gas. }
\label{fig_app_temp_map}
\end{figure}

 We first initialize the system in a thermal state at temperature $T$ via classical Monte Carlo, and then calculate its energy $E= \la H_0 \ra$ using the Hamiltonian in Eq. \ref{eq_hubbard}.
By varying the temperature of the system $T$ while keeping the total number of atoms $N$ fixed, we calculate the energy $E$ as a function of $T$.
In Fig. \ref{fig_app_temp_map} we show the temperature dependence of the energy $E$ for the pure- and quasi-2D gas that are described in Sec. \ref{sec_nonequilibrium}.

 To determine the heating, we first stir the gas with the repulsive stirring potential as described in Sec. \ref{sec_sim_method} and then after stirring calculate its energy $E$ using Eq. \ref{eq_hubbard}. We numerically invert this energy $E$ to the equilibrium temperature $T_\meq$ using the temperature dependence shown in Fig. \ref{fig_app_temp_map}. Finally, from the temperature difference between the stirred and initial system, the heating $\Delta T_\meq = T_\meq - T$ is determined.

\begin{figure}
\includegraphics[width=0.95\linewidth]{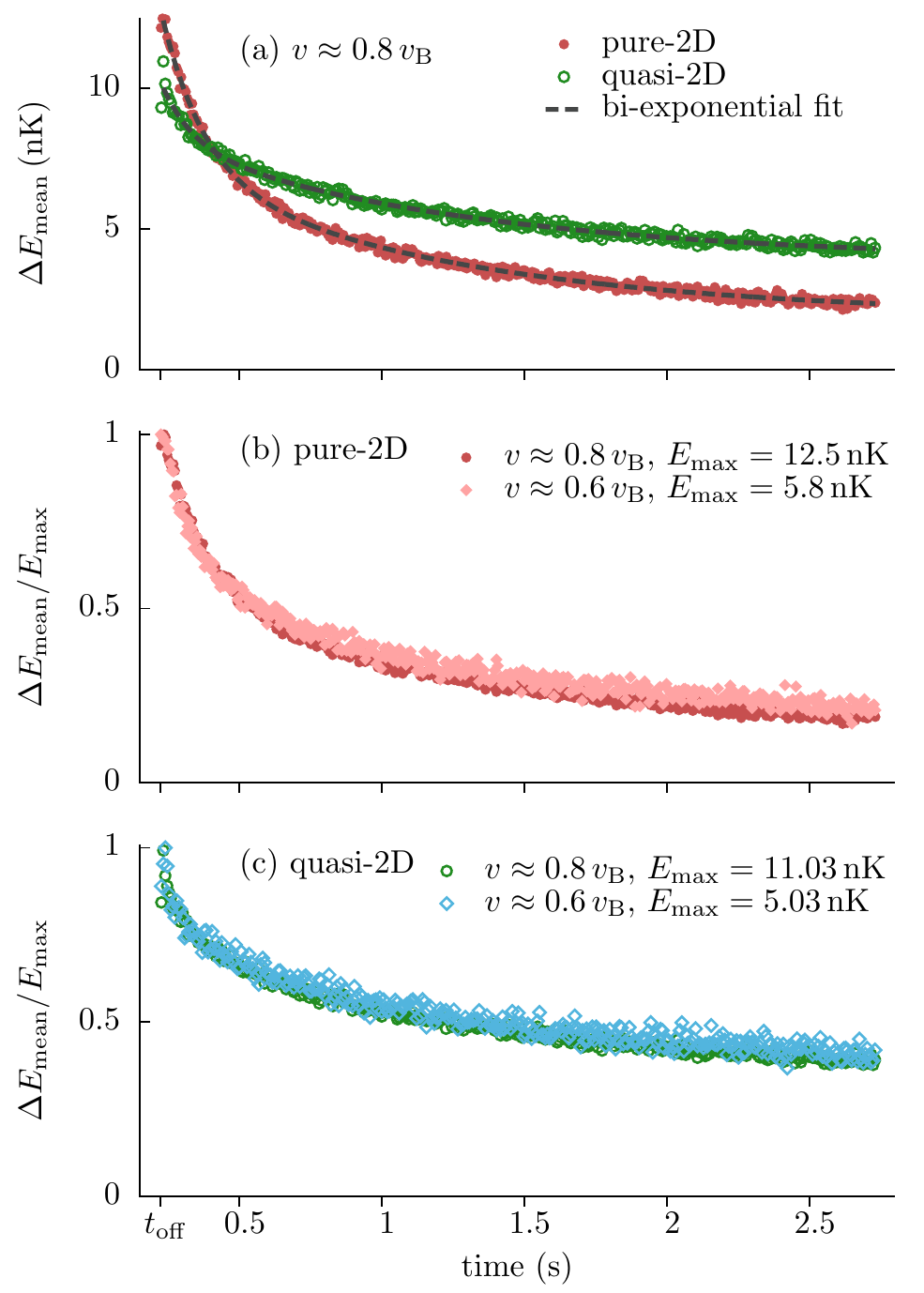}
\caption{In panel (a) we show the evolution of the mean excess energy $\Delta E_\mathrm{mean}$ for the pure- and quasi-2D gas stirred at velocity $v \approx 0.8 \, v_\mB$. We fit the energy decay with the bi-exponential function in Eq. \ref{eq_fit_decay} to determine the decay times  $\tau_{d_1},\, \tau_{d_2}$ (see text). The fitted curves are shown by the dashed lines.
We show $\Delta E_\mathrm{mean}$ normalized by its maximum mean energy $E_\mathrm{max}$ corresponding to $v \approx 0.8 \, v_\mB$ and $0.6 \, v_\mB$ for the pure- and quasi-2D gas in panels (b) and (c), respectively. }
\label{fig_app_energy_decay}
\end{figure}

\section{Relaxation dynamics}

  In this section we elaborate on the relaxation dynamics of the stirred trapped gas that we discuss in Sec. \ref{sec_nonequilibrium}. We first discuss the energy flow dynamics in Sec. \ref{app_energy_flow} and then the vortex dynamics in Sec. \ref{app_vortex_dynamics}.

\subsection{Energy-flow dynamics}\label{app_energy_flow}

  Here we elaborate on the energy flow dynamics for the pure- and quasi-2D trapped system. 
We stir both systems with the stirring potential at velocity $v \approx 0.8 \, v_\mB$. 
After stirring we calculate the excess energy $\Delta \tilde{E}_i = \bigl(E_i(t) - E_i^\meq \bigr)/n_\mmax$ as described in Sec. \ref{sec_nonequilibrium} and by averaging this energy over the superfluid region of the gas, we calculate the mean energy $\Delta E_{\mathrm{mean}}$. We show the evolution of $\Delta E_{\mathrm{mean}}$ for both systems in Fig. \ref{fig_app_energy_decay}(a). 
$\Delta E_{\mathrm{mean}}$ decays approximately exponentially as the excess energy $\Delta \tilde{E}_i$ within the superfluid region outflows to the thermal cloud. We quantify the energy decay time using the fitting function in Eq. \ref{eq_fit_decay}. 
From the fit, we determine the decay times $\tau_{d_1},\, \tau_{d_2} \approx 0.13,\, 0.98\, \ms$ and  $0.08,\, 1.02\, \ms$ for pure- and quasi-2D gas, respectively.
These decay times are similar to the ones that we determine from the vortex decay in Sec. \ref{sec_vortex}.
In addition to the stirring at $v \approx 0.8 \, v_\mB$, we show $\Delta E_{\mathrm{mean}}$ corresponding to stirring at $v \approx 0.6 \, v_\mB$ for the pure- and quasi-2D gas in Fig. \ref{fig_app_energy_decay}(b) and (c), respectively.

\begin{figure}[H]
\includegraphics[width=0.95\linewidth]{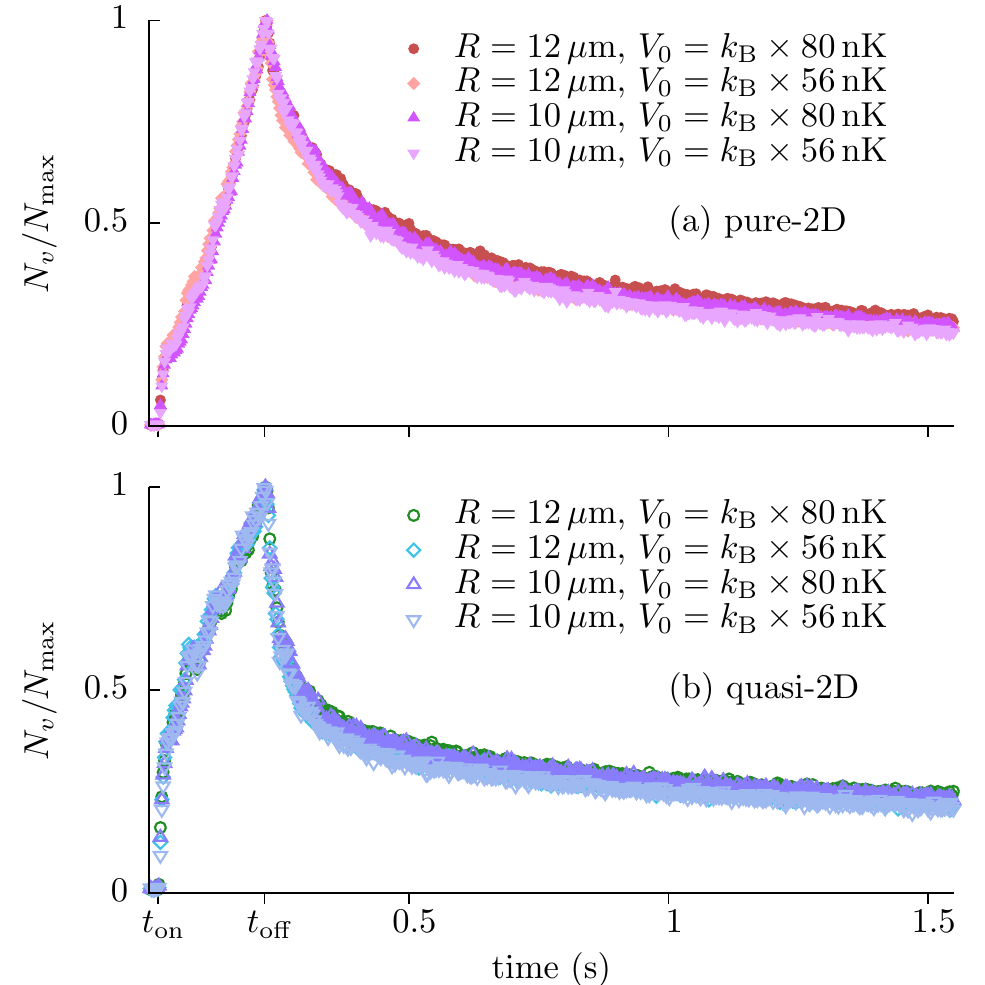}
\caption{We show the averaged vortex number $N_v$ normalized by its maximum vortex number$N_\mathrm{max}$ for the different stirring radii $R$ and strengths $V_0$, for the pure- and quasi-2D gas in panels (a) and (b), respectively. }
\label{fig_app_vortex_decay}
\end{figure}

\subsection{Vortex relaxation}\label{app_vortex_dynamics}

  Next, we turn to the vortex relaxation of the stirred gas. We stir the pure- and quasi-2D gas using different stirring radii $R$ and stirrer strengths $V_0$. We calculate the total number of vortices within the superfluid region of the cloud, as described in Sec. \ref{sec_vortex}, and average it over $128$ realizations. We show the averaged vortex number $N_v$ normalized by its maximum vortex number $N_\mathrm{max}$ as a function of time $t$ for the pure- and quasi-2D gas in Figs.  \ref{fig_app_vortex_decay}(a) and  \ref{fig_app_vortex_decay}(b), respectively. For the pure-2D gas the relaxation of vortices is slower than that for the quasi-2D gas, as shown in Sec. \ref{sec_vortex}.


\appendix


\begin{thebibliography}{999}
\bibitem{remark1} For the pure- and quasi-2D gas the border of the superfluid region with the final equilibrium temperature is expected to be at $r_\mBKT = 15.3\, \mu \mm$ and $16.4\, \mu \mm$, respectively.
\end{thebibliography}
\end{document}